# Quasi bound states in the continuum in terahertz free-standing metal complementary periodic cross-shaped resonators

Dejun Liu[1, 2], Feng Wu[3], Lin Chen[4], Feng Liu[1,2]

[1]Department of Physics, Shanghai Normal University, Shanghai 200234, China
[2]Key Laboratory for Submillimeter Astrophysics, Shanghai Normal University, Shanghai 200234, China
[3]School of Optoelectronic Engineering, Guangdong Polytechnic Normal University, Guangzhou 510665, China
[4]Shanghai Key Lab of Modern Optical System, University of Shanghai for Science and Technology, Shanghai 200093, China

Corresponding authors: dejunliu1990@gmail.com; fengwu@gpnu.edu.cn; fliu@shnu.edu.cn

We numerically and experimentally achieve quasi-bound states in the continuums (BICs) with high-Q factors in the free-standing metal complementary periodic cross-shaped resonators (CPCRs) at terahertz (THz) frequencies. Such induced quasi-BICs arises from the breaking of the mirror symmetry of CPCRs. By properly tuning the asymmetric factor, the measured Q factor of quasi-BIC can reach 102, which is lower than the simulated Q factor of 166 due to the limited system resolutions. We also simulate the electric field magnitude and vector distributions at the quasi-BICs, where the out-phase alignment between the electric dipoles is found. The sharp quasi-BICs realized in this thin free-standing metal structure may immediately boost the performance of filters and sensors in terahertz wave manipulation or biomolecular sensing.

In recent years, a kind of special resonant mode with infinite quality (Q) factor called bound states in the continuum (BICs) have attracted researchers' interest [1]. As the parameter (e.g. structural parameter or incident angle) slightly deviates from that at BICs, BICs turn into quasi-BICs with ultra-high Q factors [2-7]. Owing to its strong resonance, quasi-BICs can be intensively utilized in ultra-sensitive sensing [7,8], ultra-narrowband filtering [9-11], and enhancing nonlinear effects [12-13]. Particularly, realizing BICs within terahertz (THz) range becomes a hot topic due to THz waves' non-ionization and low photon energy (4.1 meV at 1 THz) [14-19]. To this end, different micro-structures such as metasurfaces [14], photonic crystals [15], and resonators [17] have been proposed, accelerating the based components toward the practical applications. Previously, different types of BICs have been reported in metasurfaces [6,13,16,18]. By suitable in-plane symmetry breaking, BICs can be produced in metasurfaces [16,18]. For example, BICs emerge in metasurfaces formed by arrays of detuned resonant dipolar dimers. This resonance evolves continuously from a Fano resonance into a symmetry-protected BIC as the dipole detuning vanishes [16]. In addition, a quasi-BIC resonance in metasurfaces consisted of split ring resonators has been applied for THz sensing, which enables sensing of a 7 nm thick analyte [18]. Compared to the quasi-BIC based metamaterials or metasurfaces, free-standing metal structures that do not require any holder or substrate show high flexibility [20-25]. Furthermore, free-standing metal structures have no additional loss from other materials. Terahertz free-standing metal structures based on surface plasmon polaritons (SPPs) have been investigated frequently in past decades [22-27]. Owing to the SPPs, electromagnetic waves are efficiently trapped at the metal edges through the interaction with the free electrons of the metallic surfaces [22]. Typical structures such as metal hole arrays [22-24], metal grooves arrays [26], and metal slits [27] have been reported recently. However, oblique waves are applied to excite the sharp resonance in metal hole arrays (MHAs), restricting the MHAs for practical applications [25]. Although the Ohmic losses of metal contribute to the total losses, a resonator based on metal structures with high-Q BIC resonances and strong SPPs is highly desirable in THz applications, particularly in ultra-sensitive sensing.

In this letter, we numerically and experimentally achieve sharp resonances with high-Q factors in the free-standing metal complementary periodic cross-shaped resonators (CPCRs) at THz frequencies. Such sharp resonance with Q of 102 can be termed as quasi-BICs, arising from the breaking of the mirror symmetry of CPCRs, which could be further improved by reducing the structural asymmetry factor. In order to deep understand the resonant characteristic, electric field magnitude and vector distributions are also simulated. The achievement of high-Q factors in the THz region indicates that such thin metal structures are especially interesting for practical THz applications in wave manipulation and biomolecular sensing.

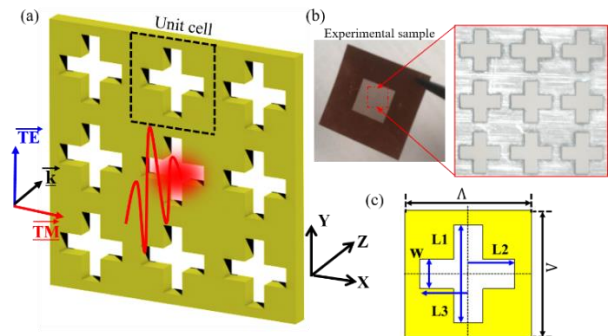

FIG. 1. (a) Configuration of complementary periodic cross-shaped resonators (CPCRs). (b) Photos of experimental samples. (c) Unit cell of CPCRs.

Figure 1 shows the configuration of the metal structures, which consists of complementary periodic cross-shaped resonators (CPCRs) with subwavelength scales. The structure is free-standing, i.e. without using a dielectric substrate. The material of the 50 μm-thick metal sheet is selected as copper with the conductivity of 5.96E7 S/m. The perforated films are usually optically thin, i.e., much smaller than the wavelength (>0.25 mm). As a result, this thin metal structure can achieve strong SPPs coupling between top and bottom metal surfaces [22]. The structure has four sharp metal corners (90 degree), which is different from conventional circular-hole arrays [22-25]. The structural parameters of the unit cell are as follows: Λ=260 µm, L1=200 µm, L3=100 µm, and w=60 µm. L2 is a variable parameter, ranging from 80 to 120 µm. When L2 equals to 100 µm, the in-plane mirror symmetry along y-axis is preserved. By changing L2, the in-plane mirror symmetry along y-axis can be broken. The simulation results are performed by FDTD (Finite-difference time-domain) methods, where the periodic boundary conditions are applied to the x- and y-directions and absorbing boundary conditions are applied to the z-direction. A plane wave is normal incident in the structure, where the field of TM and TE modes is parallel to the metal surface.

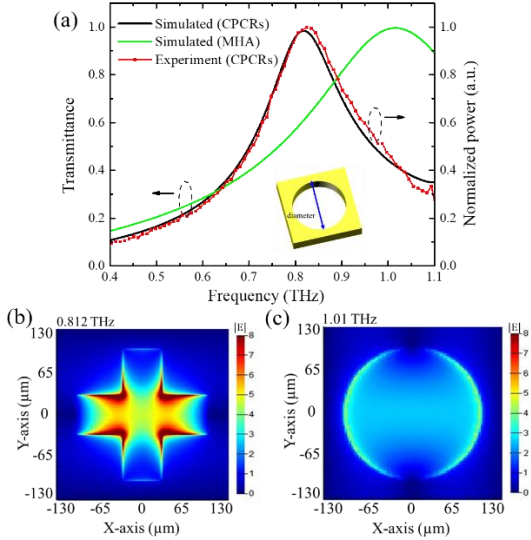

FIG. 2. (a) Transmittance spectra of complementary periodic cross-shaped resonators (CPCRs) and metal hole array (MHA). (b-c) Electric field distribution of 0.812 and 1.01 THz for CPCRs and MHA, where the color bar value is the normalization of the local electric field to the incident electric field.

Figure 2(a) shows the simulated and experimental transmittance spectra of CPCRs using FDTD methods and terahertz time-domain spectroscopy (THz-TDS) measurements, where the experimental results show excellent agreement with simulated ones. To further understand the transmission properties of CPCRs, a metal hole array (MHA) with diameter of 200 µm is investigated. The equation of resonance frequencies can be expressed as $f_r=c/((i^2+j^2)^{1/2}nd)$ $(i,j=\pm 0, \pm 1...)$, where $n$ is the effective refractive index in the hole, $d$ is the diameter of holes [22]. From this equation, it is noted that the spectral peak frequency is tightly associated with the hole diameter. As shown by the green line in Fig. 2(a), a resonant peak is observed at 1.01 THz, which corresponds to the lowest order SPP mode ($f_r$=1.05 THz). For CPCRs, the structural parameter L2 is 100 µm, which means that the CPCR unit cell does not break in-plane symmetry. Hence, a characteristic dip/peak pair (Fano-type lineshape) in the spectra is not observed in the CPCRs because the symmetry of the structure is preserved, which corresponds to the ideal BIC [1, 6, 10]. The lowest resonant peak of CPCRs occurs at 0.812 THz, corresponding to a wavelength of 0.370 mm. Compared to the MHA with the same period, the introduced four corners in CPCRs enhance the SPPs coupling and result in a lower resonant frequency. We simulated the electric field distribution of 0.812 THz for CRCPs, as shown in Fig. 2 (b). It is clear that the strong SPPs field of 0.812 THz is concentrated at the metal corner tips and decays gradually from the mesh/air interface. In contrast to the field distribution of MHA at 1.01 THz (Fig. 2(c)), the CRCPs show enlarged field covered regions, thus it would be expected as a THz sensor.

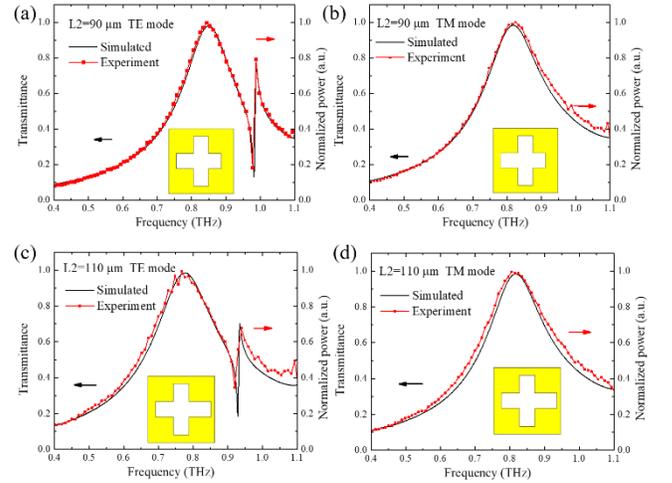

FIG. 3 Transmittance spectra of CPCRs with L2= 90 µm for TE (a) and TM (b) modes incidences. Transmittance spectra of CPCRs with L2= 110 µm for TE (c) and TM (d) modes incidences.

In order to achieve quasi-BIC with high-Q factor, now we break the mirror symmetry of the structure. The CPCRs is composed of four arms, whose mirror symmetry can be broken by increasing or reducing the length of one arm. Figure 3 shows the simulated and experimental transmittance spectra for different lengths of L2. As shown in Fig. 3(a), for the TE modes, a sharp resonant dip at 0.974 THz is occurred, where the simulated results are in excellent agreement with our experimental findings. In the case of the sharp resonator, the resonance characteristics strongly depend on the asymmetry parameter, defined as δ=|(L3-L2)|/(L2+L3) [17]. Here, the asymmetry parameter is 0.052 when L2 is 90 µm. The sharp dip can be termed as a quasi-BIC Fano resonance, which shows a high Q of 153, where the Q factor can be calculated by the equation of $f_0/\Delta f$ in which $f_0$ and $\Delta f$ is resonance frequency at the dip and the resonant width ($\Delta f = f_{peak}-f_{dip}$), respectively [28]. When L2 equals to 110 µm, corresponding to an asymmetry parameter of 0.047, the induced resonance at 0.923 THz has a Q factor of 166, which is higher than that of 90 µm because of the lower asymmetry parameter (Fig. 3(c)). Noted that the measurement of the largest Q values is limited by sweep resolution (9.0 GHz) and signal-to-noise ratio. Hence, the measured Q factor is 102, which is lower than the simulated one of 166. For the TM modes, as shown in Figs. 3(b) and (d), the quasi-BIC does not occur since the mirror symmetry along x-axis is unbroken [13].

To further understand the BIC induced by TE modes, we simulated the transmittance spectra map of CPCRs (copper) for various L2, ranging from 80 to 120 µm (Fig. 4(a)). The perturbation theory precisely and efficiently predicts the resonance frequency shift due to the small change of the structure [29]. It can be seen that the resonance frequency slightly decreases as L2 increases. For example, when the L2 is 80 µm, the resonance locates at about 1.0 THz, which is higher than 0.974 THz when L2 is 90 µm. Specifically, when L2 locates in the region of 98 to 102 µm but except for 100 µm, although the mirror symmetry is broken, the quasi-BIC resonance is not obvious. Different from ideal PEC, the Ohmic losses in copper makes the weaker resonance. In addition, the reduction of edge charge density at the wedge also contributes to the decrease of the Q factor [30].

As L2 increases from 105 μm to 120 μm, the frequency of resonances decreases from 0.931 THz to 0.907 THz. Meanwhile, the linewidth of quasi-BIC resonances increases with the increasing L2. Thus, the linewidths and frequencies of both quasi-BICs reveal strong dependence on the degree of structural asymmetry. Figure 4 (b) depicts the simulated spectra of the ideal CPCRs composed of PEC and realistic metallic (copper) with varying structural parameters L2. It is obviously seen that PEC CPCRs show sharper resonances than that of copper CPCRs owing to their zero Ohmic losses.

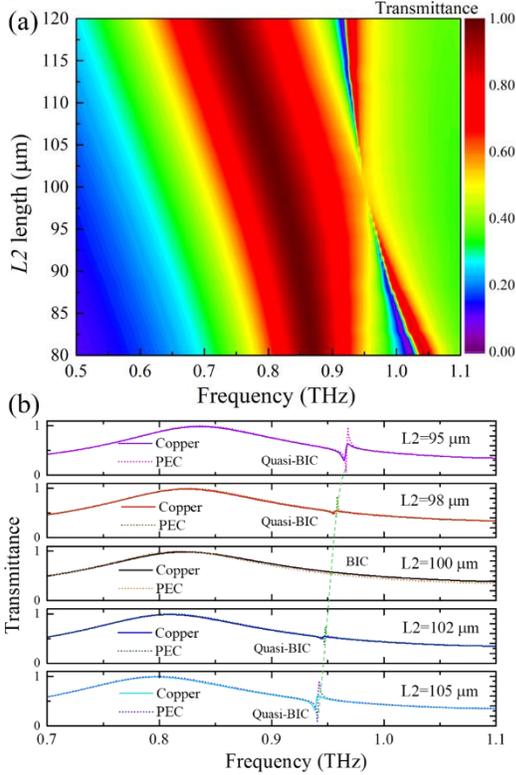

FIG. 4. (a) Simulated transmittance spectra map of CPCRs for various L2. (b) Simulated spectra of the ideal CPCRs composed of PEC and realistic metallic (copper) with varying structural parameters L2.

We summarize the Q factor of BIC resonances for PEC and copper CPCRs with a varying ΔL (L2-L3), as shown in Fig. 5(a). As ΔL gradually approaches to near zero, the Q factor increases rapidly. For instance, for the PEC CPCRs, the Q factor reaches about 1164 in the case of ΔL = ±1 μm. However, for the copper CPCRs, owing to the Ohmic loss, the corresponding Q factor is about 600 and lower than that of PEC CPCRs. When ΔL is equal to zero, the resonance width vanishes completely and the Q factor becomes infinite. This case corresponds to BIC. The relation of Q factor for the PEC CPCR with $1/\delta^2$ is also given in Fig. 5 (b). For PEC CPCRs, the Q factor increases with the increasing $1/\delta^2$. Noted that the Q factor is almost proportional to $1/\delta^2$, which is similar to Refs. [6] and [7].

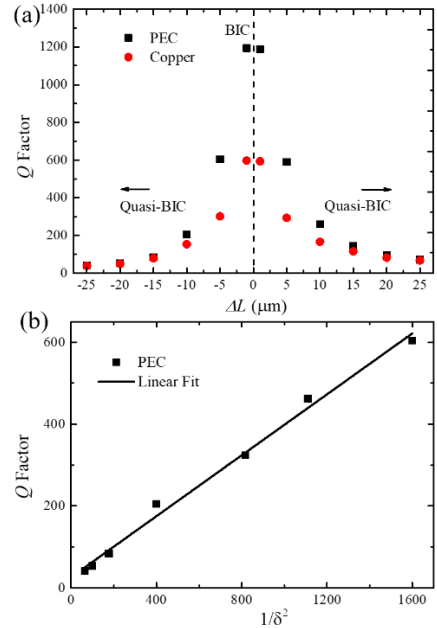

FIG. 5. (a) Dependence of Q factor on ΔL. (b) Linear relationship between Q factor and $1/\delta^2$. The black line is a linear fitting line.

When the L2 equals to 90 μm and 110 μm, the induced quasi-BICs respectively occur at 0.974 and 0.923 THz. In this section, two sharp resonances induced by the difference lengths of L2 will be investigated by electric field magnitude and vector distributions. The simulated electric field magnitude and vector distributions are shown in Fig. 6, where the cut plane is the surface of CPCRs. To better understand resonant characteristic, the lowest resonant peaks at 0.854 (L2=90 μm) and 0.780 THz (L2=110 μm) are selected as examples to explore. The mirror symmetry breaking, as discussed in the previous studies, is expected to affect the intensity of the fields [16-17]. As shown in Fig. 6(a), the strong fields at 0.854 THz transfer to the left side of CPCRs in comparison to that of 0.812 THz when L2=100 μm (see Fig. 2(b)). It means that the symmetry-broken structure causes an asymmetrical field distribution. Similar phenomena can also be found in Fig. 6 (c) when the L2 is 110 μm, the strong field at 0.780 THz is located at the right side of CPCRs. Figures 6(b) and (d) show the electric field magnitude distributions of quasi-BIC resonances for L2=90 and 110 μm. At 0.974 THz for L2=90 μm, the enhanced fields are confined at the metal corner tips, where the maximum value of the color bar is 20. Interestingly, the field in the center of the aperture is very weak, which is different from the conventional resonance shown in Fig. 6(a). A similar electric field distribution can be observed at 0.923 THz when the L2 is 110 μm. Figures 6(e-h) show the corresponding electric field vector distributions for theses frequencies. At 0.854 THz, when the L2 is 90 μm, the electric dipoles locate at the gap between the metals and its alignment are in-phase (or symmetry). Interestingly, the out-phase alignment between the electric dipoles not only in horizontal but also vertical can be found at the quasi-BIC resonance of 0.974 THz, which is different from the resonance at 0.854 THz. It means that the dipole symmetry is broken in terms of the orientation, which allows coupling of the mode to incident TE waves for quasi-BIC resonances [17]. As a result, strong coupling between the horizontal and vertical dipoles are occurs because of the symmetry-broken effect [6]. A similar dipole alignment can also be found in Figs. 6(g-h) when L2 is 110 μm. As we can see, at 0.780 THz, the electric dipoles align in-phase but the direction is contrary with that of 0.854 THz. For instance, the electric field direction is downward at 0.780 THz, but at 0.854 THz is upward directions. The difference between 0.974 THz and 0.923 THz is more obvious, where the L2 length is 90 and 110 μm, respectively. In Fig. 6(h), the out-phase alignment between adjacent dipoles can also be observed. However, the dipole direction of 0.923 THz is contrary to that of the resonance at 0.974 THz because of the different phase.

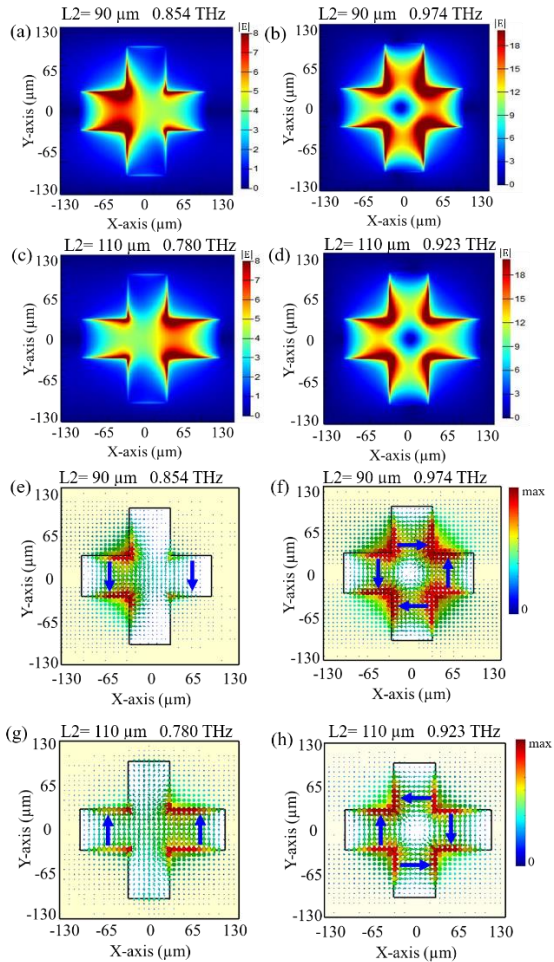

FIG. 6. (a-b) Electric field magnitude distributions and (e-f) vector distributions (e-f) for 0.854 and 0.974 THz when L2 is 90 μm. (c-d) Electric field magnitude distributions and (g-h) vector distributions for 0.780 and 0.923 THz when L2 is 110 μm. The color bar value in (a-d) is the normalization of the local electric field to the incident electric field.

In conclusion, by breaking the in-plane mirror symmetry of the structure, we numerically and experimentally demonstrate quasi-BICs with high-Q factors in the free-standing metal CPCRs at THz frequencies. Compared to the MHA with the same thickness, CPCRs achieves stronger localized fields at the metal surface. The measured Q factor of the quasi-BIC at 0.923 THz reaches 102 when the arm length of the CPCRs equals to L2=110 μm. From the simulated electric field magnitude and vector distributions, the out-phase alignment between the electric dipoles not only in horizontal but also in vertical can be found at quasi-BICs. These results demonstrated that such quasi-BICs in pure metal structures could be exploited as filters and sensors for practical applications in terahertz wave manipulation and biomolecular sensing.